\documentclass[11pt,a4paper]{article}
\usepackage{amsmath}
\usepackage[utf8]{inputenc}
\usepackage[T1]{fontenc}
\usepackage{fixltx2e}
\usepackage{longtable}
\usepackage{float}
\usepackage{wrapfig}
\usepackage{soul}
\usepackage{textcomp}
\usepackage{marvosym}
\usepackage{wasysym}
\usepackage{latexsym}
\usepackage{amssymb}
\usepackage[pdftex]{graphicx}
\usepackage[margin=1in]{geometry}
\usepackage{sidecap}  
\usepackage[table]{xcolor}  
\usepackage{lineno,xcolor}
\linenumbers
\modulolinenumbers[2]
\usepackage{mathtools}  
\mathtoolsset{showonlyrefs}  
\usepackage[square,compress,numbers]{natbib}
\newcommand{\kmer}{$k$-mer}
\newcommand{\kmers}{$k$-mers}

\title{}
\author{Nicholas Putnam}
\date{\today}

\begin{document}

\section*{}
\section*{Joint assembly and genetic mapping of the Atlantic horseshoe crab genome reveals ancient whole genome duplication.}
\label{sec-2}

\vspace{0.3in}

Carlos Nossa\textsuperscript{1}, 
Paul Havlak\textsuperscript{1}, 
Jia-Xing Yue\textsuperscript{1}, 
Jie Lv\textsuperscript{1},  
Kim Vincent\textsuperscript{1},  
H Jane Brockmann\textsuperscript{3}, 
Nicholas H Putnam\textsuperscript{1,2}

\vspace{0.3in}
\begin{small}
(1) Department of Ecology and Evolutionary Biology, and (2) Department of
Biochemistry and Cell Biology, Rice University, P.O. Box 1892, Houston
TX 77251-1892; (3)
Department of Biology, P.O.B. 11-8525, University of Florida,
Gainesville, FL 32611-8525.  
\vspace{0.3in}

Address for correspondence: 
\vspace{0.1in}

nputnam@rice.edu
\vspace{0.1in}

Nicholas Putnam

Ecology and Evolutionary Biology -- MS 170

Rice University

P.O. Box 1892

Houston, TX 77251-1892

\end{small}

\vspace{0.3in}

Keyword:  Genotyping-by-Sequencing (GBS)

Keyword:  Genetic linkage mapping

Keyword:  Genome evolution 

Keywork:  Limulus polyphemus

\vspace{0.3in}

\newpage
\section*{Abstract}
\label{sec-3}

Horseshoe crabs are marine arthropods with a fossil record extending
  back approximately 450 million years.
  They exhibit remarkable morphological stability over their long
  evolutionary history, retaining a number of ancestral arthropod
  traits, and are often cited as
  examples of ``living fossils.''  
As arthropods, they belong to
  the \emph{Ecdysozoa}, an ancient super-phylum whose sequenced genomes
  (including insects and nematodes) have thus far shown more divergence
  from the ancestral pattern of eumetazoan genome organization than
  cnidarians, deuterostomes, and lophotrochozoans. However,
  much of ecdysozoan diversity remains unrepresented in comparative
  genomic analyses.  
Here we use a new strategy of combined \emph{de novo} assembly and
  genetic mapping to examine the chromosome-scale genome organization
  of the Atlantic
  horseshoe crab \emph{Limulus polyphemus}.  We constructed a genetic
  linkage map of this 2.7 Gbp genome by sequencing the nuclear DNA of 34
  wild-collected, full-sibling embryos and their parents at a mean
  redundancy of 1.1x per sample.  The map includes 84,307 sequence
  markers and 5,775 candidate conserved protein coding
  genes. Comparison to other metazoan genomes shows that 
 the \emph{L. polyphemus} genome preserves ancestral bilaterian linkage
  groups, and that
  a common ancestor of modern horseshoe crabs underwent one or more ancient
  whole genome duplications (WGDs) \texttildelow 300 MYA, followed by extensive
  chromosome fusion.

\newpage
\section*{Introduction}
\label{sec-4}

  Comparative analysis of genome sequences from diverse metazoans has
  revealed much about their evolution over hundreds of millions of
  years.  The discovery of extensive gene homology across large
  evolutionary distances has allowed researchers to track chromsome
  rearrangements and whole genome duplications.  The resulting value
  of whole chromosome sequences presents a challenge for existing
  whole genome shotgun (WGS) assembly strategies.

  Whole genome duplication events were long 
 suspected\citep{ohno_evolution_1970}, but only the availability of
 genome sequences has allowed confirmation of them in fungal,
 vertebrate, plant, and ciliate 
lineages\citep{wolfe_molecular_1997,mclysaght_extensive_2002,simillion_hidden_2002,aury_global_2006}.
 In contrast, when only a few chordate, insect, and nematode genomes were
 available, conservation of gene linkage (i.e. synteny) and gene order
 were observed only between closely-related species, and consequently
 were not expected to be conserved between phyla.  As more metazoan
 genomes have been sequenced, it has become clear that long-range
 linkage has been conserved over long time scales in many lineages.

Sequencing the genomes of representatives of chordate, mollusk, annelid,
 cnidarian, placozoan and sponge clades has identified 17 or 18
 ancestral linkage groups
 (ALGs)\cite{putnam_sea_2007,putnam_amphioxus_2008,srivastava_trichoplax_2008,srivastava_amphimedon_2010,simakov_insights_2012}.
 Each of these ALGs consists of a set of ancestral genes whose
 descendants share conserved synteny in
 multiple sequenced genomes.  These ALGs have been interpreted to
 correspond to ancestral metazoan chromosomes, and correlations
 between inferred rates of gene movement between ALGs across the
 metazoan tree suggest that these ancestral linkage relationships are
 conserved through the action of selective constraints on a subset of
 genes\cite{lv_constraints_2011}.

  The relatively small number of genomes from anciently
  distinct metazoan lineages and the fragmented nature of draft
  genome assemblies still limit both the search for ancient whole genome
  duplications and the power of the data to constrain
  models of chromosome-scale genome structure evolution.
  While WGS sequencing technology and assembly
  methods are active areas of research and technological development,
  and have improved at a dramatic pace in recent years, high quality \emph{de   novo} assembly of large, complex metazoan genomes remains a difficult
  and resource-intensive problem.  Without genetic or physical maps, or
  reliance on a high-quality reference genome of a closely-related
  species, WGS sequencing projects still typically produce assemblies
  containing thousands of scaffolds, hundreds of scaffolds incorrectly
  joining sequence from different chromosomes, or
  both\citep{earl_assemblathon_2011}.

  Next-generation sequencing has greatly reduced the cost of
  constructing high density genetic maps by eliminating the need to
  develop and genotype polymorphic markers
  individually\cite{davey_genome-wide_2011}.  This has been achieved
  either by focusing sequence coverage within or adjacent to genomic
  regions of distinct biochemical character, such as restriction sites with
  RAD-seq and related methods\cite{altshuler_snp_2000,baird_rapid_2008}, or by combining
  information across regions using a reference genome
  sequence\cite{huang_high-throughput_2009,andolfatto_multiplexed_2011}.
  While RAD-seq is applicable to organisms lacking a reference genome
  assembly, it is not directly applicable to comparisons of genome
  organization across long evolutionary time spans because such
  comparisons rely on the identification of homologous sequence markers
  (typically protein-coding genes), which typically have only a small overlap with
  the restriction-associated markers.

  Here we present a genotype-by-sequencing method for constructing a
  high-density genetic map using low-coverage, low-cost, whole genome
  sequencing data from the offspring of a wild cross.  In this joint
  assembly and mapping (JAM) approach the traditionally independent and
  sequential steps of genome assembly, polymorphic marker
  identification, and genetic map construction are combined.  Existing
  assemblers expect lower densities of sequence polymorphism, deeper
  coverage, greater computer memory, or more aggressive quality trimming
  that decrease sequence coverage\cite{chapman_meraculous:_2011,butler_allpaths:_2008,gnerre_high-quality_2011}.
  Our current implementation focuses on conservative assembly of short
  scaffolds sufficient for map construction, but our results suggest
  that further integration of genetic mapping information within whole
  genome shotgun assembly methods can be a cost effective
  way to produce assemblies of large,  complex genomes with
  chromosome-scale contiguity.

We have applied this approach to produce a genetic
  map of the genome of the Atlantic horseshoe crab, \emph{Limulus   polyphemus}.  
  Horseshoe crabs are marine arthropods with a fossil record extending
  back \texttildelow 450 million years\citep{rudkin_horseshoe_2009}.
  They exhibit remarkable morphological stability over their long
  evolutionary history, retaining a number of ancestral arthropod
  traits\citep{fisher_xiphosurida:_1984}, and are often cited as
  examples of ``living fossils.''   
  \emph{L. polyphemus} has
  a large genome about 90\% the size of the human genome. It is an
  important species from ecological, commercial, and conservation
  perspectives\citep{berkson_horseshoe_1999}, that has been used as a
  model system for research in behavioral ecology, physiology and
  development\cite{shuster_jr._american_2003}.  The map and SNP
  markers described here will be a resource for the \emph{L. polyphemus}
  genome project, research in horseshoe crab population biology, and
  comparisons of metazoan genome organization.  By anchoring protein
  coding genes to this map, we are able to extend anaylsis of
  ancestral linkage groups and whole genome duplications to the
  chelicerate lineage.
\section*{Results}
\label{sec-5}
\subsection*{Assembly and Mapping}
\label{sec-5-1}

The JAM method is designed to produce a combined assembly of
polymorphic sequences, tagged by genomic regions with at most one SNP
per \emph{k}-mer window (Methods).  Starting with genomic reads from a
mating pair of adult \emph{L. polyphemus} and 34 offspring (\texttildelow 100 bp reads
on \texttildelow 300 bp inserts), we tabulated high-quality 23-mers occurring in
two sequencing runs (and thus also in at least two individuals).

\begin{figure}[htb]
 \centering
 \includegraphics[width=.9\linewidth]{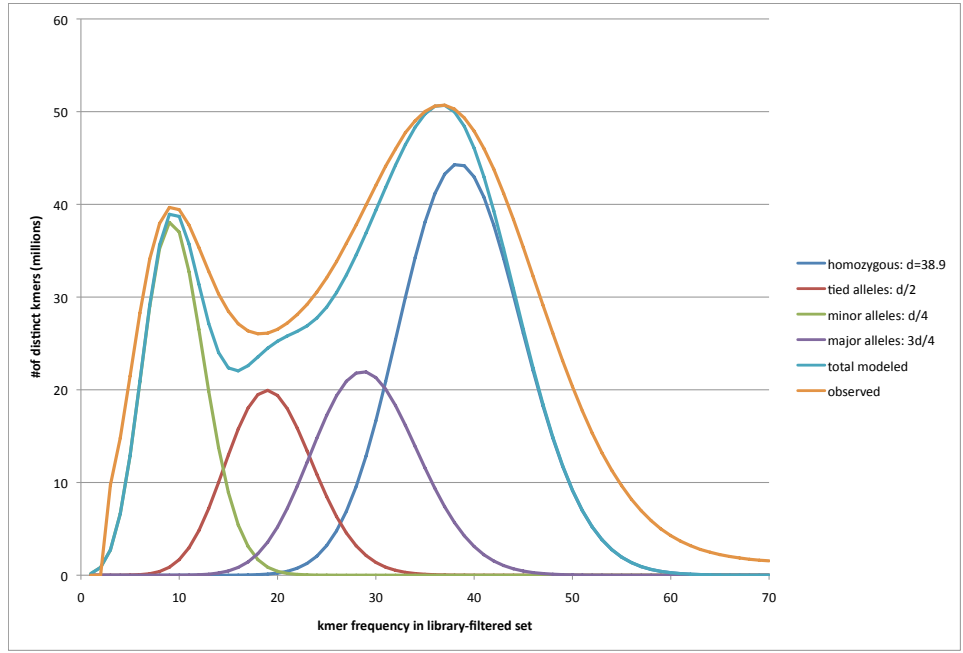}
 \caption{\label{fig:Lp_decomp}Fitting Poisson distributions to Limulus 23mer frequencies (filtered as described below)}
 \end{figure}

Fitting Poisson models for unduplicated sequence to the frequencies of these
filtered 23-mers suggests that 1.1 billion genomic loci are unique at this resolution.
Of these \emph{k}-loci, 63 percent are modeled as homozygous, 
27 percent as paired major-minor alleles, 
and the remaining 10 percent as tied allele pairs (Figure \ref{fig:Lp_decomp}).
The corresponding SNPs, if always at least 23 bases apart, 
would be 1.6 percent of bases in the \emph{k}-loci.
Dividing the total number of filtered 23-mers by the modeled
homozygous depth of coverage $d = 38.9$ yields an estimated genome
size of 2.74 billion bases, consistent with the measured DNA content of 2.8
pg (978 Mb $\simeq$ 1 pg DNA)\citep{gregory_animal_2012}.

We categorize specific 23-mers by their edit distances to others,
having no neighbors within a single base substitution
(unique tags) or with a single mutually unique neighbor (``SNPmer pair'' tags).
A subset of these, including SNPmer pairs for approximately 7.9 million SNPs,
constitute the tags used for contigging and scaffolding.
The SNP-mer pairs account for about 45 percent 
of the modeled fraction of alleles,
the others missed from similarity to other sequences 
(e.g. due to repeats) or distance from each other 
(because of indels or multiple SNPs per 23-mer).

Chaining these 23-mers together (see Methods) produces an initial 6.6
million contigs, 3.9 million of which are linkable by paired reads for
scaffolding. Applying Bambus\citep{pop_hierarchical_2004} produces 944 thousand scaffolds spanning
1.3 million bases (Table \ref{table:ContigStats}). These scaffolds serve as
markers incorporating multiple \emph{k}-loci, including SNPmer pairs
used to identify haplotypes.

\begin{table}[htb]
\caption{\emph{K}-mer contig and scaffold statistics} \label{table:ContigStats}
\begin{center}
\begin{tabular}{lllrr}
 Assembled             &  count      &  total (bp)     &  avg. span (bp)  &  n50 span (bp)  \\
\hline
 \emph{k}-mer contigs  &  6,614,434  &  1,240,275,515  &             188  &            418  \\
 linkable contigs      &  3,925,844  &  1,137,576,911  &             290  &            460  \\
 initial scaffolds     &  944,246    &  1,261,263,172  &            1336  &           3047  \\
 reference scaffolds   &  944,246    &  1,295,334,515  &            1372  &           2930  \\
 reference bases       &             &  1,131,458,744  &            1198  &           2553  \\
\end{tabular}
\end{center}
\end{table}

After assembly, the mean density of single nucleotide polymorphisms (SNPs) across the
four haplotypes in assembled regions was estimated based on read
re-alignments to be 7.6 per
thousand bases.  We jointly inferred the phases of these SNPs and
segregation pattern (offspring genotypes) in the mapping cross for
each marker in a maximum likelihood framework (Methods).  We focused on the
91,320 markers with at least 18 inferred bi-allelic SNPs for
constructing the linkage map.  These markers grouped into 1,908
high-confidence map bins (\emph{i.e.} unique segregation patterns, assumed
to correspond to loci in the genome uninterrupted by meiotic
recombination in the cross\citep[][]{van_os_construction_2006}).  Map
bins fell into 32 linkage groups, close to the 26
pairs (2N=52) previously found in a cytogenetic analysis of two
chromosome spreads\citep{iwasaki_karyology_1988}.  20 map bins were
removed for having inconsistent positions in the maternal and paternal
maps, and 12 were singletons.

To estimate the frequency of incorrect genotype calls as a function of
the log likelihood difference between the called and alternative
genotype (genotype confidence score), including contributions from
uncertainty in SNP-mer identification, assembly, and sampling noise,
we carried out a simulation of the library pooling and sequencing, \kmer\ assembly and genotype inference protocols, using the sequenced \emph{Ciona intestinalis} genome as a starting point.

\begin{figure}[htb]
\centering
\includegraphics[width=.9\linewidth]{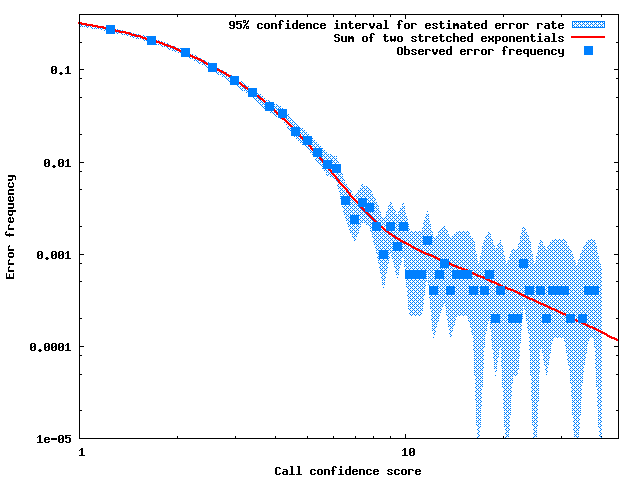}
\caption{\label{fig:errors}Genotype call error rate as a function of call confidence score for bins of 10,000 calls in simulated \emph{Ciona intestinalis} genome data.  The stippled blue region shows 95\% confidence intervals of the Bayesian posterior probability distribution of the underlying error rate computed from the Beta distribution Beta$(n_e+1,n_c-n_e+1)$ conjugate the assumed binomial distribution of observed errors, where $n_e$ and $n_c$ are the number of errors and number of calls in each bin respectively.}
\end{figure}

In the simulated \emph{C. intestinalis} data set (Methods), a single stretched
exponential distribution provided a good fit to the frequency of genotype calling
errors as a function of the call confidence score for scores up to
6, or down to error frequency of about 1\%.  The observed error
frequency declined more slowly for higher confidence scores.  The
minimum $\chi^2$ fit used for estimating the genotyping error rate in the
\emph{L. polyphemus} map bins 
was $p_e(s)=a_1 e^{ -s^{c_1} / b_1 } + a_2 e^{-s^{c_2} / b_2}$, with
parameter values $a_1 = 0.49, b_1 = 2.08, c_1 = 1.26, a_2 = 5.47, b_2 = 0.17, c_2 = 0.16$
 (Figure \ref{fig:errors}). 

Applying this model to the \emph{L. polyphemus} marker genome calls, we
estimated that the genotype calling error rate in the map bin
representative markers was 0.0099.  51\% of adjacent map bin pairs are
separated by a single inferred recombination event in the cross, and 94\% are
separated by three or fewer recombinants in each parent.

Of the 91,320 markers with at least 18 putative SNPs, 84,307 (92\%)
were assigned to their closest map bins with a threshold of $p <
10^{-6}$ (Methods), for an estimated genome-wide average density of one mapped
sequence marker every 32 kb.  A mean of 45 markers were mapped to each map
bin, and the number of markers mapped was used to estimate the
relative physical size of map bins.  46\% of the scaffolds with 12-17 SNPs
could be placed with the same threshold, for an additional 32,688 markers, or
one marker every 23 kb.

The total length of the scaffolds assigned to map bins was 411 Mb, and
they contained 2.67 million bi-allelic SNPs assigned a phase with a
posterior probability of at least 0.99.  Of these, 72\% were inferred
to be unique to one of the four parental chromosomes.  This is close
to the 74\% predicted under the finite sites neutral coalescent
model given the observed SNP density\citep{cartwright_family-based_2012}.

\begin{figure}[htb]
\centering
\includegraphics[width=.9\linewidth]{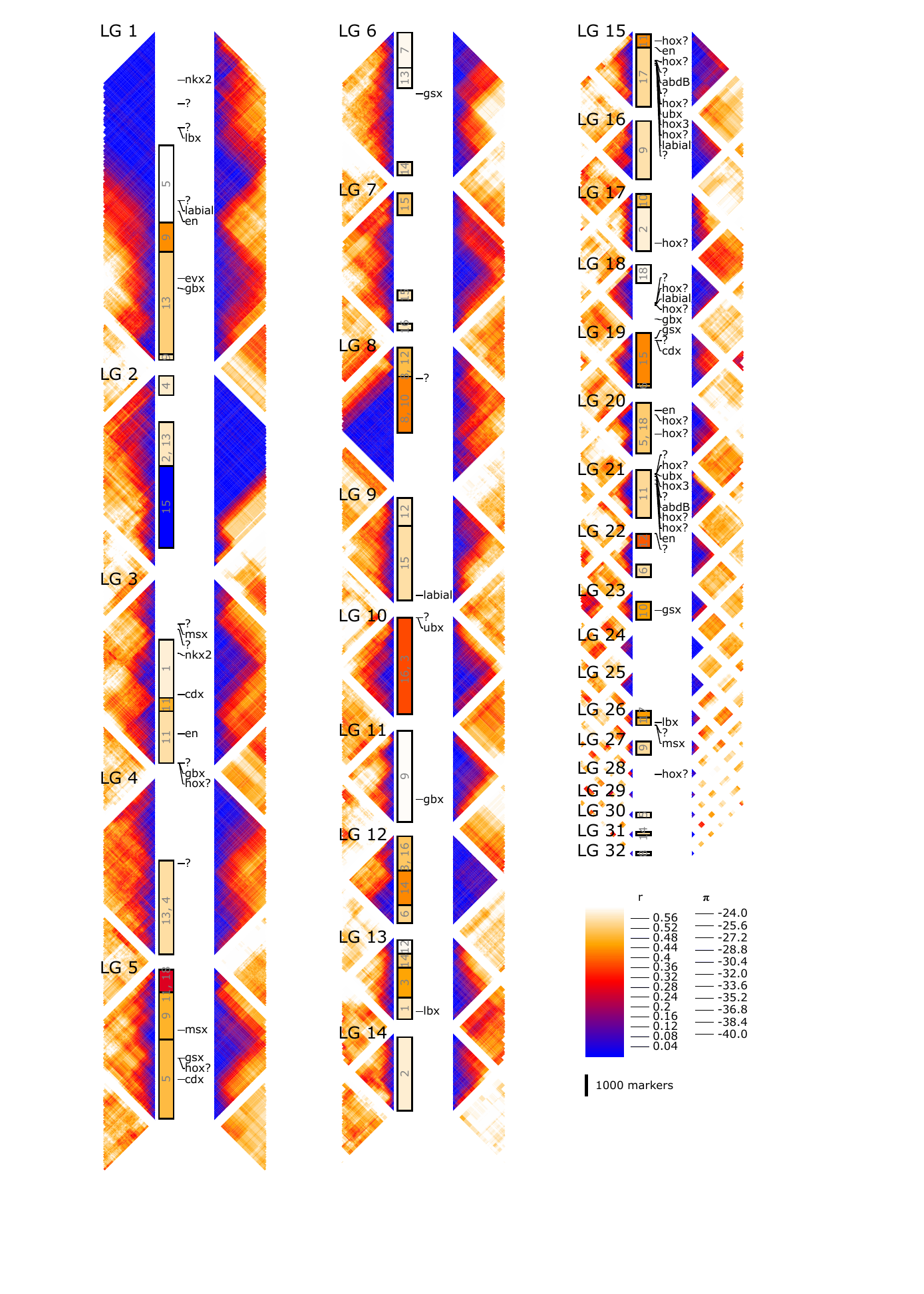}
\caption{\label{fig:matrix}Each of the numbered blocks represents one of the 32 linkage groups of the \emph{L. polyphemus} genetic map, and is composed of four columns: Two bands of the triangular matrices in which the color scale indicates the fraction of samples showing recombination between pairs of markers; maternal recombination frequency is shown on the left, paternal on the right.  A column labeled ``ALG'' indicates segments of significant (p<0.05 in Fisher's Exact Test, after Bonferroni correction for multiple tests) conservation of gene content with ancestral bilaterian linage groups.  The column labeled HOX shows the map positions and types of predicted homeobox transcription factor genes.  The two color scales are for: recombination frequency between pairs of markers and log p-value for enrichment in gene content with ancestral linkage groups.}
\end{figure}
\subsection*{Sequence composition and recombination rate}
\label{sec-5-2}

In the scaffolds longer than 1kb (mean length 2.9 kb), the $G/C$ base
content was 33.3 $\pm$ 2.8 percent, and the local relative frequency
of CpG dinucleotides was bimodally distributed, with about 30\% of sequences exhibiting depletion of
CpG.  TpG and CpA dinucleotides were over-represented on average
and their local densities
negatively correlated with CpG density, suggesting ongoing germ-line
CpG methylation for a fraction of the genome\citep{bird_dna_1980}.

The mean maternal and paternal recombination rates were estimated to
be 1.28 and 0.76 centimorgans per megabase respectively, consistent
with expectation based on genome size\citep{lynch_origins_2006}.
  We did not observe evidence of segregation distortion
for any map bins.
Estimated
local recombination rates in the two parents are correlated across the
genome with and $r^2$ = 7.1\% ;
p<1e-29, and the mean local recombination rate is correlated with
local SNP density,  $r^2$ = 9.7\% ; p<1e-40.
\subsection*{Ancestral linkage group conservation}
\label{sec-5-3}

\begin{figure}[htb]
\centering
\includegraphics[height=5in, trim=0cm 16cm 12cm 0cm, clip=true, angle=90]{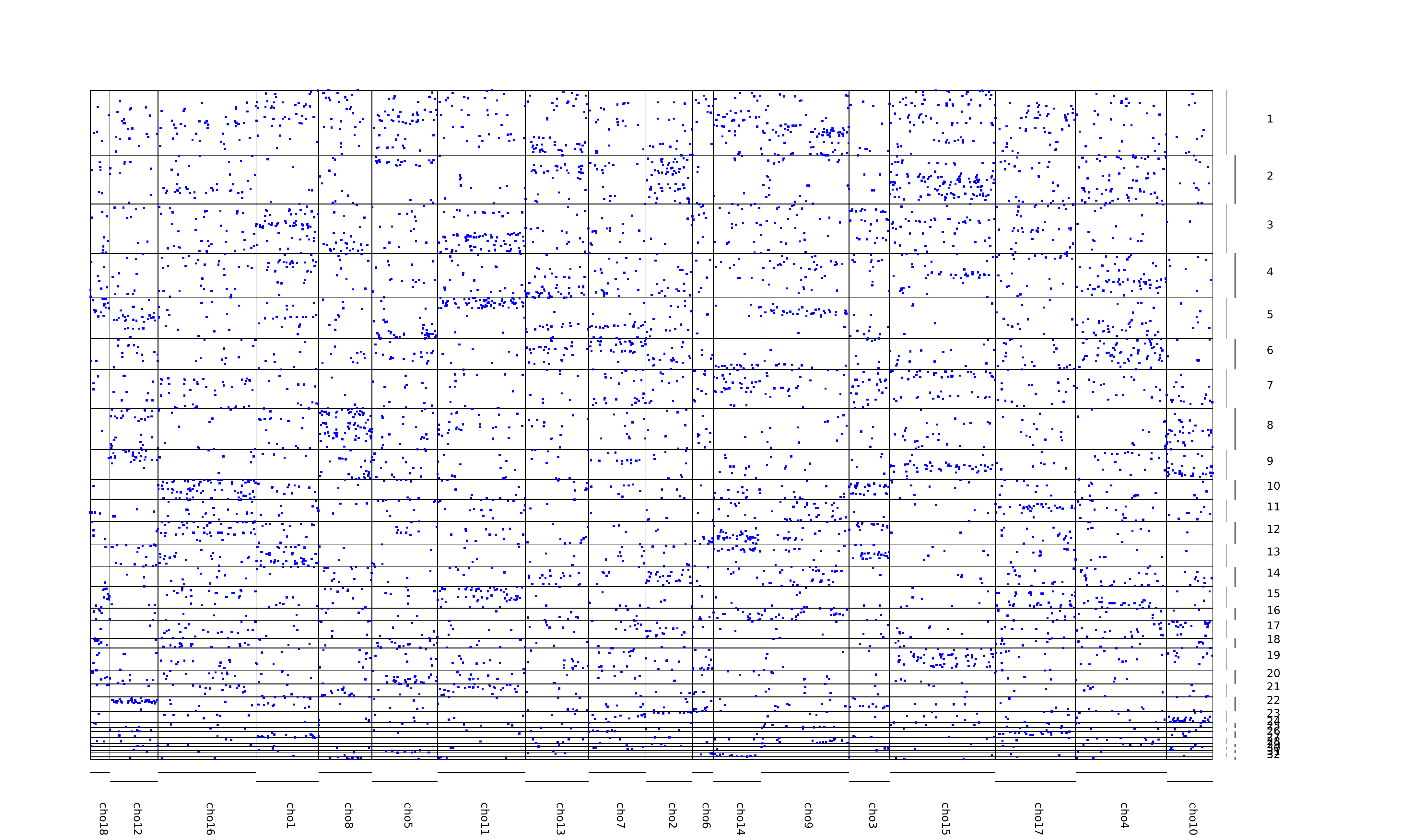}
\caption{\label{fig:dotplot}Limulus-Human macro-synteny dot plot.  Blue points indicate the position of human genes in reconstructed ancestral chordate ALGs (vertical displacement) and their candidate orthologs in the 30 \emph{L. polyphemus} linkage groups (horizontal displacement).}
\end{figure}

\begin{figure}[htb]
\centering
\includegraphics[height=5in, trim=0cm 16cm 12cm 0cm, clip=true, angle=90]{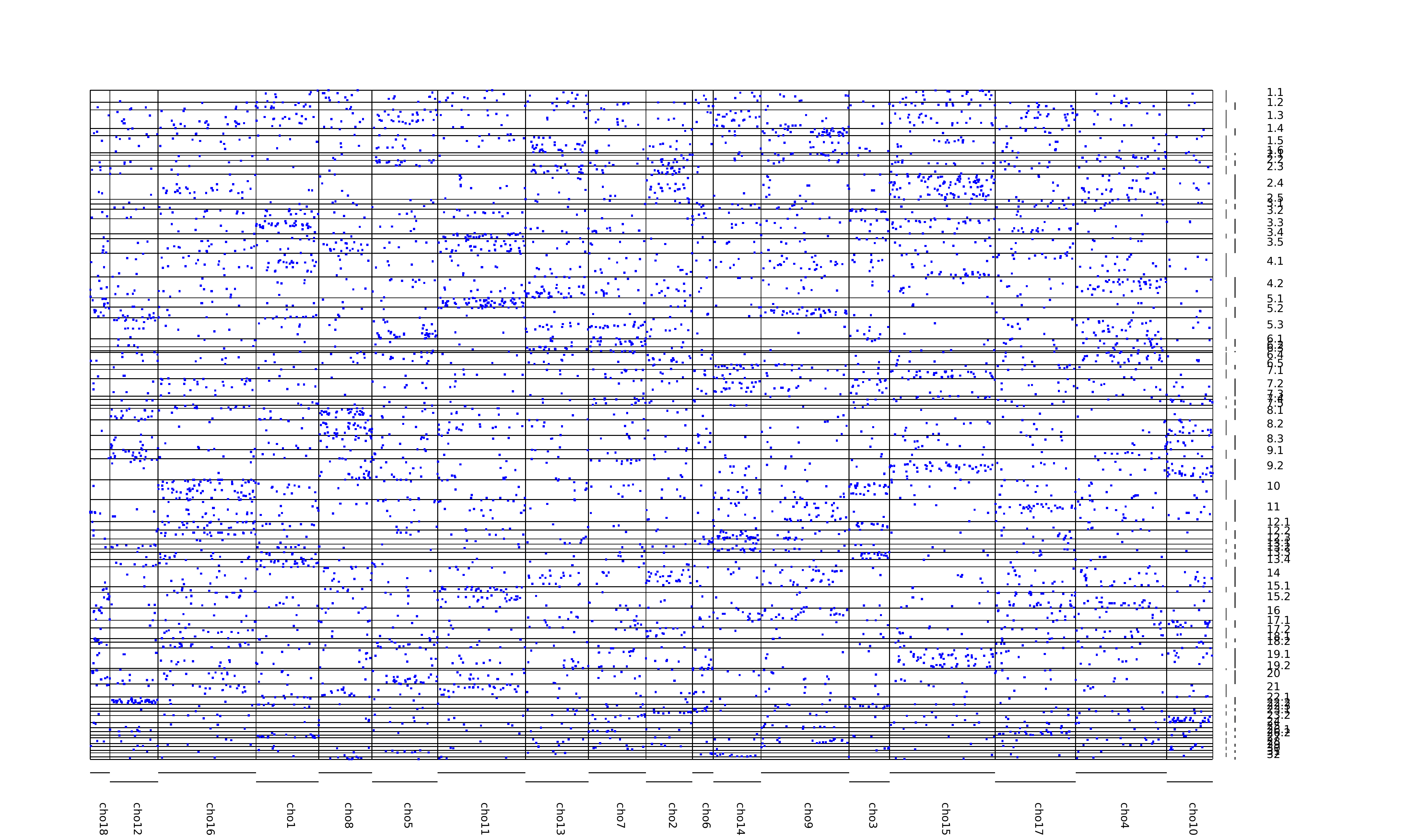}
\caption{\label{fig:dotplot2}Limulus-Human macro-synteny dot plot as in Figure \ref{fig:dotplot}, showing breaks introduced by hidden Markov model segmentation of the linkage groups as vertical lines.}
\end{figure}

34,942 scaffolds have significant sequence conservation with 10,399
predicted proteins of the tick \emph{Ixodes scapularis}\citep{vectorbase_ixodes_2008}. 6,246 of these hits formed
reciprocal best pairs, of which 5,775 (92\%) could be placed on the
linkage map at a threshold of p<$10^{-6}$.  These were used as
conserved markers for comparisons of genome organization.  When
linkage groups were divided into 108 non-overlapping bins of 1,000
markers, 52 had significant (p<0.05, after Bonferroni correction for
1,944 pairwise tests) enrichment in shared orthologs (or ``hit'') with at
least one of eighteen ancestral chordate linkage
groups\citep{putnam_amphioxus_2008}.  A hidden Markov model
segmentation algorithm\cite{putnam_sea_2007} identified 40 breakpoints
in ALG composition in the linkage groups.  72\% of the genome is
spanned by 53 intervening segments that hit one or (for eight of them)
two ALGs (Figures \ref{fig:dotplot} and \ref{fig:dotplot2}).  Each of
the eighteen ancestral ALGs has at least one hit among the 45
segments with a unique hit to the ALGs.

\subsection*{Whole genome duplications}
\label{sec-5-4}

  Whole genome duplication (WGD), or polyploidization is a rare but
  dramatic genetic mutation event which doubles the size of a genome and
  creates a redundant pair of copies from every gene.  Because it
  creates redundant copies of genes for entire biochemical pathways and
  genetic networks, it has been proposed that it creates unique raw
  material for the evolution of novel biological functions and increased
  complexity.
\subsection*{Homeobox gene clusters}
\label{sec-5-5}

Homeobox genes encode a large family of transcription factors involved
in diverse embryonic patterning and structure formation processes of
eukaryotes. As a particular subfamily of homeobox genes, the Hox
cluster is known to control metazoan body patterning along the
anterior-posterior axis.  We identified 155 scaffolds with significant
homology to predicted chelicerate homeobox gene sequences in public
databases.  We classified these sequences into homeobox subfamilies (Methods)
and placed them on the map by best hit. Two large clusters of Hox
genes are found on linkage group (LG) 15 and LG 21, each containing multiple members
of the anterior, central and posterior classes.  There are also two
parahox cluster homologs, each with three homeobox genes: \emph{gsx} and
\emph{cdx} orthologs and a third homeobox gene not confidently assigned
to a subfamily in our analysis (LG 5 and LG 19). There are two smaller
clusters containing multiple hox genes (LG 18 and LG 20), and clusters
of other homeobox genes, including members of the \emph{msx}, \emph{lbx}, \emph{nk},
\emph{evx}, and \emph{gbx} families (Figure 1).
\subsection*{Genomic distribution of paralogous genes}
\label{sec-5-6}

WGD creates many pairs of duplicate genes or ``paralogs''.  The
distinctive features of these genes have been used to infer WGD events
in fungal, vertebrate and plant genomes\citep{wolfe_molecular_1997,mclysaght_extensive_2002,simillion_hidden_2002}.
We examined the genomic distribution of 2,716 pairs of candidate
paralogous gene markers in \emph{L. polyphemus} for signatures of WGD.  In
45\% of these pairs both markers mapped to the same chromosome,
compared to 5.3 $\pm$ 0.5 \% in 1000 datasets with randomly-permuted
paralogous gene identities.  The mapped positions of pairs within the
chromosomes were highly correlated (average $r^2$  = 0.81, and exceeding
0.95 for 8 of the large chromosomes; Figure \ref{fig:paralogdots}),
suggesting that many of the pairs represent recent tandem gene
duplicates or single genes fragmented across multiple markers.  In the following, these
same-chromosome paralogs are referred to as ``tandem'' duplicates.

Inter-chromosomal duplicates are clustered into conserved paralogous
micro-synteny blocks (or ``paralogons''\cite{mclysaght_extensive_2002}): there are 25 pairs of
loci, each with at least six ($m_p$ =6) paralog pairs clustered with
a maximum gap (max-gap) of 300 markers between adjacent paralogs in
each cluster.  These clusters span 25,044 markers, or 30\% of the map, after removing
redundancy from paralogons with overlapping footprints.  In 1000
datasets with randomly-permuted paralogous gene identities, the
maximum number of such clusters observed was 11; the mean and standard
deviation were 3.9 $\pm$ 1.0.  The observed clustering into paralogons
was greater than that in the randomized datasets over a broad range of
choices of max-gap and $m_p$.  For example, for max-gap=100, $m_p$ =3
there are 52 clusters vs. 3.5 $\pm$ 1.9, range 0-10; for max-gap=500,
$m_p$ = 9 there are 12 vs. 2.9 $\pm$ 1.7, range 0-9.  Because of the
large proportion of apparent tandem gene duplicates (45\%), this
randomization scheme increases the number of inter-chromosomal paralog
pairs relative to the data, making it a conservative significance test
for inter-chromosomal paralog clustering.  When genes with tandem
duplicates are excluded from the randomizations, the observed number
of clusters is greater than the maximum observed in 1000
randomizations for all the combinations of max-gap in the set
(100, 200, 300, 400, 500, 600) and $m_p$ in the set (3, 4, 5, 6, 7, 8,
9).  23
max-gap=600, $m_p$ = 7 clusters span 59\% of the map, compared to respective mean
number and map coverage of 3.3 $\pm$ 1.8, and 13 $\pm$ 6 \%
in these randomizations.  

Among the marker pairs mapping to different chromosomes, we find a
significant excess of pairs relating segments derived from the same
ancestral linkage group (ALG) relative to randomization controls (247
pairs vs 102 $\pm$ 11, p<0.001 in randomizations of all genes; 202 vs
46 $\pm$ 7 when genes in tandem duplicates are excluded). This pattern is consistent with the
creation of these segments by duplication (rather than fission).

The max-gap clusters have a significant amount of overlap among their
footprints.  For example, the footprints of the max-gap=600, $m_p$ = 7 clusters had a
total length of 72,072 markers, but a net footprint after redundnacy
removal of 49,545 markers.  We examined the relationships among the
paralogons for evidence of successive rounds of duplication.  We
considered a graph in which nodes correspond to merged, non-redundant
paralogon footprint regions.  Nodes are connected with edges if a
max-gap cluster connects the two nodes.  The average clustering
coefficient of this graph is equal to the probability that footprints
$a$ and $c$ share a max-gap cluster, given that there are edges
$(a,b)$ and $(b,c)$ in the graph.  We compared the clustering
coefficients to those found in random Erdős-Rényi graphs with the same
number of nodes and edge probability as the observed graph.  We found
that the observed data shows significantly more clustering than these
random graphs for a wide range of choices of max-gap and $m_p$.  For
example, for max-gap 600, $m_p$ =7, the average clustering coefficient is
0.19, while 10,000 random graphs had coefficients of 0.034 $\pm$ 0.042,
p=0.0039.  
\subsection*{Age distribution of paralogous genes}
\label{sec-5-7}

Because WGD events create many paralogs at the same time, they leave
characteristic peaks in the age distribution of paralogous genes.  In
\emph{L. polyphemus} the distribution shows peaks centered at 0.71 and 1.34
substitutions per synonymous site (K$_s$), values within the
approximately linear response range of K$_s$ estimates to WGD age\citep{vanneste_inference_2013} (Figure \ref{fig:ks}).
For comparison, the synonymous site divergence between an Asian
horseshoe crab species \emph{Tachypleus tridentatus} and \emph{L. polyphemus}
has a mode of 0.35.  The common ancestor of these species has been
estimated to have lived 114-154 MYA, coincident with the opening of
the Atlantic ocean\citep{obst_molecular_2012}, suggesting a WGD event
230-310 MYA, and possibly an older one 450-600 MYA.
\section*{Discussion}
\label{sec-6}

Our results demonstrate that a low cost, combined approach to whole
genome sequencing and genetic mapping can be used to efficiently
create a very high density genetic recombination map for a non-model
organism with a large genome.  Because the approach uses genome-wide
sequencing, a large number of sequence markers can be anchored to the
map, allowing comparisons of genome organization at the chromosome
scale over very large evolutionary divergences.  The identification of
chromosomal segments with significant gene composition homology to
each of the chordate ALGs shows that the predominance of fusion and
mixing of ancestral linkage groups previously observed in analyzed
ecdysozoan genomes\cite{simakov_insights_2012} is not ancestral to or
universal in the clade.
 
The map allows quantiative characterization of other features of
chromosome-scale organization, such as the correlation between local
recombination rate and polymorphism levels.  Similar positive
correlations between local recombination rate and polymorphism level
have been observed in other metazoans including
humans\cite{begun_levels_1992,cutter_selection_2003,nachman_single_2001}
and plants\citep{stephan_dna_1998,roselius_relationship_2005}.  Future
comparisons to more closely related chelicerates will allow tests to
distinguish whether these rates are positively correlated with
inter-specific divergence, consistent with a neutral process of
correlated mutation and recombination
rates\cite{hellmann_neutral_2003}.  Alternatively, the association
could be explained by hitchhiking and background selection\cite{andolfatto_regions_2001}.

The existence of duplicated hox and parahox clusters on four
different chromosomes is highly suggestive of multiple whole genome
duplication.  Hox clusters have not been found in duplicate copies
except in vertebrates where they have been created by whole genome
duplication, and have only rarely been subsequently lost.

The enrichment of inter-chromosomal paralog pairs in segments of the
same ALG origin is consistent with their creation by duplication
(rather than fission), although because small-scale duplication is
biased toward local (tandem) duplication, fission of segments could
also leave behind an enrichment of paralogs.  Such a mechanism,
however, would not create the observed organization of 
paralogs, \emph{i.e.} their clustering into ``paralogons''.  The fact
that these paralogons span a large portion of the map (59\%) suggests
that it was a whole genome duplication, rather than segmental
duplications that have rise to the pattern.

The double-peaked shape of the distribution of synonymous site
divergence between pairs of paralogs, combined with the existence of
two small clusters of HOX genes in addition to the two complete HOX
clusters suggests that there may have been two rounds of whole genome
duplication in the horseshoe crab lineage.

WGDs preceded major species radiations in vertebrates, angiosperms and
teleost fish and the importance of their role in evolution is the
subject of long-running debate\citep{wolfe_molecular_1997,
mclysaght_extensive_2002, simillion_hidden_2002, ohno_evolution_1970}.
The discovery of whole genome duplication in an invertebrate, and
during horseshoe crabs' long and famously conservative evolutionary
history suggests that such events may have been more common than
previously assumed in metazoan evolution, and that while they may have
provided raw material for adaptive evolution in some cases, they are
not evolutionary drivers.
\section*{Methods}
\label{sec-7}
\subsection*{Joint assembly and mapping (JAM) overview}
\label{sec-7-1}

Barcoded genomic DNA libraries were created, pooled, and sequenced in
four lanes on the Illumina HiSeq2000 platform for a mating pair of \emph{L.  polyphemus} and 34 offspring (Methods).

  The JAM method proceeds through three major phases: 1. The frequencies
  of DNA sub-sequences of fixed length $k$ (\kmers) are profiled to
  characterize the quality, uniqueness, polymorphisms and repetition in
  genomic reads, using software we developed building on work from the
  Atlas assembler\citep{havlak_atlas_2004}.  Allelic pairs of \kmers\ 
  representing alternate forms of single nucleotide polymorphisms (SNPs)
  are identified and tracked through the subsequent steps.  2.
  Contigs are assembled on a graph of unique \kmers\  and paired SNP \kmers\, sampled to reduce memory usage, then ordered and oriented
  using the Bambus
  scaffolder\citep{pop_hierarchical_2004,treangen_next_2011}.   Each
  multi-SNP scaffold is treated as a single marker for the linkage
  mapping steps. 3.  The paired SNP \kmers\  (SNPmers) in each scaffold are combined with the read,
  mate-pair, and parent- or offspring-library associations of their
  alleles for haplotype phasing and construction of a high density
  genetic linkage map.
\subsection*{Sampling and sequencing}
\label{sec-7-2}

Tissue samples were collected from the third walking leg of a
monandrous pair of horseshoe crabs nesting at high tide on the beach
at Seahorse Key, an island along the west coast of north Florida, on
27 March 2010.  The eggs laid by this pair were collected 6 hr later
when the tide had receded and reared in plastic dishes as previously
described\citep{johnson_costs_2010}.  Trilobite larvae hatched from
the eggs 4 weeks later. Tissue samples and larvae were preserved in
RNALater.  Genomic DNA purification and library construction were carried out
using Qiagen DNAEasy, Illumina TruSeq and Nextera kits, following
manufacturers' protocols. Barcoded samples were pooled and sequenced 
on on the Illumina HiSeq2000 platform.

Limulus larvae were processed as follows; each larva, suspended in 100 $\mu\text{L}$ of RNAlater and 
stored at -80 °C in a 1.5 mL Eppendorf tube, was thawed on ice, after which RNAlater was 
removed. DNA was extracted using the Qiagen DNAEasy kit per manufacturer's protocols. 
DNA was quantified using picogreen DNA quantitation kit. To prepare TruSeq libraries, 
DNA was first purified another time using zymo genomic DNA clean columns per manufacturer's protocols. 
Adult \emph{L. polyphemus} DNA was prepared as above, but using claw tissue rather than whole larvae.  
All DNA extracts were tested by gel electrophoresis to ensure DNA was not degraded. 
TruSeq libraries were prepared at University of Georgia's Georgia Genomics Facility. 
1-5 micrograms of sample DNA was subjected to fragmentation using Covaris sonicator. 
Fragmented DNA was then used for library construction using Illumina TruSeq library prep kits. 
Libraries were pooled together in equimolar amounts (for 10 larvae) and used for sequencing. 
For samples 11-34, Library prep was switched from TruSeq to Nextera kits. 
Nextera library preparation was performed according to manufacturer's protocol. 
The Nextera library product was quantified by picogreen, and fragment size distribution was checked by using Lonza flash gel, 
to ensure that fragment size distribution was between 300-1000 bp. 
Sample libraries were pooled in equimolar concentrations and sent for sequencing. 
Library pools were sequenced on the Illumina HiSeq2000 platform at Medical College of Wisconsin Sequencing Service Core Facility.
\subsection*{\kmer\  decomposition}
\label{sec-7-3}

We determine a lower bound on the \kmer\  size long enough for a given
expectation of uniquenes in a random genome. While increasing \emph{k}
reduces the rate of coincidentally repeated \kmers, it also reduces
the effective depth of coverage due to untrimmed errors and edge
effects at read ends — and increases the cases of multiple SNPs per
\kmer\  locus, which are not tracked in our current software implementation.  We
can approximately model a genome-scale string \emph{G} of random nucleotides as 
$G$ samples taken with uniform probability from the space of all \kmers\  
(of size $4^k/2$ for odd \kmers\, treating reverse-complements as same; slightly
more for even \emph{k}). The Poisson distribution then gives the probability that a
location in \emph{G} has its own, unique \kmer\  (shared with no other location) as
\begin{quote}
\centering $e^{-\lambda}$, where $\lambda = G/(4k/2) = 2G/4k$.
\end{quote}
The probability of a location sharing its \emph{kmer} is then $1 - e^{-\lambda}$; thus, to 
limit the maximum rate \emph{R} of \emph{G}-locations sharing \kmers, we 
require $k \geq \lceil \log_4(-2G/\ln(1-R))\rceil$. For 
example, for a mammalian-scale genome of approximately 3 billion bases, and \emph{R} = 0.1\%, we choose $k \geq 22$. For \emph{Limulus polyphemus}, the Animal Genome Size Database\citep{gregory_animal_2012} reports an estimated haploid genome size of 2.80 pg and, as each picogram represents almost a billion nucleotide base-pairs of DNA, the mammalian-scale choice of \emph{k} applies.\citep{mullikin_phusion_2003,butler_allpaths:_2008}

This lower bound ignores chemical and biological sequence biases, so
selecting $k$ for a real genome project requires attention to error
rates, repeats tandem and interspersed, and genome size, all known
vaguely, if at all, before sequencing. Studying the \kmer\ 
distributions after sequencing can clarify these genomic properties as
we select \emph{k} to maximize the net yield of candidate \kmer\  tags,
between errors and with at most one SNP location, in sequencing reads.
We convert Illumina/Solexa FASTQ format (paying attention to the
different quality encodings of the software versions) into FASTA
format\citep{cock_sanger_2010}, masking (replacing with ‘N’) any base
with Phred-scale\citep{ewing_base-calling_1998} quality below 20, and
soft-masking (representing in lower case) other bases with quality
below 30. For initial trimming experiments, we vary these quality
thresholds as indicated below.  We store \kmers\  in hash tables with
open addressing\citep{knuth_searching_1973}, supporting odd $k <=
31$. We tally for each kmer a bit vector for presence or absence in up
to 64 libraries, and an overall count of occurrences in all libraries
(count limited to $64-2k$ bits) Where the kmer hash would be too large
for available memories, we sample the kmers using a hash-slicing
factor $S$ (must be prime). Representing each kmer as an integer in
$[0..{4^k}-1]$, slice $s$ consists of those kmers whose remainder on
division by $S$ is $s$. We can tabulate one slice for a representative
sample of $1/S$ kmers (for initial estimation of depth of coverage and
genome size) or, using $S$ independent jobs, collect information for
kmers in all slices.  Our hash tables store odd-length kmers so that
reverse-complementary sequences can be combined without the ambiguous
orientation of palindromic sequences (e.g., \texttt{ATCGAT}).

After selecting \emph{k} as described above and making a full tabulation of
kmer counts and bit vectors, we filter out kmers not expected to
represent genomic sequence.  \kmers\  were required to have three copies
in the total sequence set, with at least one copy in the initial run
and one in the second run. This was partly to filter out incomplete
adapter sequences, which can be difficult to trim but which were
different in the two runs.

Extending methods developed for the Atlas assembler\citep{havlak_atlas_2004} to heterozygous sequences,
Figure \ref{fig:Lp_decomp} gives a rough decomposition of the \emph{k}-mer
frequency distribution for 23-mers with quality $\geq 20$,
minimizing the square of the residuals of \emph{k}-mer counts on frequencies 3
through 70 while not exceeding the observed counts. 
Four linked distributions model fractions of the genome as 
monoallelic or biallelic: homozygous regions with $d = 38.9$-fold coverage
(dark blue), minor alleles covered at $d/4$ (green), tied alleles at
$d/2$ (red) and major alleles at $3d/4$ (purple). This fit
is robust enough to confirm the abundance of major-minor allele pairs 
(27 percent of \emph{k}-loci, vs. 10 percent for tied alleles),
with the broader peaks in the data than in the fitted curves
consistent with less uniform sampling (for example, varying coverage
of parents and offspring). 
The Poisson decomposition suggests a density of polymorphisms of 1.2\%
in major-minor allele pairs, based on dividing the modeled number of
such sequenced pairs by \emph{k} (assuming most polymorphisms are SNPs 
spaced at least \emph{k} bases apart), by \emph{d} (the estimated depth of
sequencing) and by the estimated genome size of 2.74 billion bases.
\subsection*{SNPmer identification  \label{snpmer}}
\label{sec-7-4}

The filtered kmer counts, computed in parallel, are loaded into a hash
table with additional fields to track kmers that are uniquely within
one mismatch of each other. Because this step analyzes all (non-error)
\kmers\  in one table, this requires a single large-memory processor (on the
order of 32 GiB).

For each \kmer, we check all its $3k$ one-substitution neighbors,
The \kmers\  are partitioned each into one of three categories:
\emph{unique}: having no edit-neighbors within one
substitution;  \emph{ambiguous}: having either multiple one-substitution
neighbors, or one neighbor that has multiple neighbors; or
\emph{partnered}: uniquely pairable with exactly one other
\kmer\  differing by one substitution, such \emph{kmers} also known from now on as SNPmers
or SNPmer pairs. For each SNPmer, we save the position of the
substitution, a bitmask for the change (transition, complement, or 
non-complement transversion), 
and whether the canonical form of the partner in the
table has the same sense or is reverse complemented with respect to
this \kmer. 

Only partnered and unique \kmers\  will be further tracked. While this limited 
method cannot identify \kmers\  for genomic SNP and non-SNP locations with complete
confidence, false pairing or missed pairing should have limited effects on 
subsequent analysis. 
False pairing, due to coincidental similarities or repeats, will
combine nodes of the \kmer\  graph (see below) and cause possible misassembly: for
our purposes, noise in the scaffolding, haplotype phasing, and linkage
analysis. Missed pairing can happend from indel polymorphisms, SNPs separated
by fewer than $k-1$ positions, failure to sequence minor alleles, or 
ambiguity due to too many similar \kmers. Ambiguously non-unique \kmers\  will
be skipped over (reducing connectivity of the \kmer\  graph if there are too
many in a row). While allelic \kmers\  misidentified as unique will cause 
more forks in the \kmer\  graph, those from major-alleles 
will still be chained together with flanking unique sequence, provided
that major allele \kmers\  have at least twice the frequency as those for the
minor allele.

Table \ref{table:KmerTypes} shows the totals and percentages of the
different kmer categories, counting each SNPmer pair as one
kmer. SNPmer pairs account for 16.3 percent of the putative
genomically unique 23mer markers; dividing by 23 gives us the fraction
of bases in those markers that are putative SNPs: 0.71 percent. 

\begin{table}[htb]
\caption{K-mer categories, counting a SNPmer pair as one kmer} \label{table:KmerTypes}
\begin{center}
\begin{tabular}{lll}
 \kmer\  type  &  \# Distinct    &  Percentage  \\
\hline
 no partner/unique           &  946,431,901    &  55.48\%     \\
 partnered/SNPmer            &  184,756,149    &  10.83\%     \\
 ambiguous                   &  574,557,296    &  33.68\%     \\
 TOTAL                       &  1,705,745,346  &              \\
\end{tabular}
\end{center}
\end{table}
\subsection*{Subset \kmer\  selection \label{subset}}
\label{sec-7-5}

To reduce the memory requirements of our \kmer\  assembly graph, we
select a roughly one-tenth subset of the \kmers\ .

In the case of a true SNP at least \emph{k-1} bases from other SNPs and gaps in error-free coverage of either allele, there will be \emph{k} covering SNPmer pairs (provided that covering \kmers\  are also uniquely pairable). By taking only SNPmer pairs with the substitution in particular positions, we can reduce the size of the graph and its redundancy. Analyzing the distribution of edit position for all the SNPmer pairs, we observe an enrichment for edits near the ends due to sequencing errors. By selecting positions 3, 12 and 21 of 23-base SNPmers, we avoid the most problematic positions and reduce this portion of \kmer\  nodes by a factor of 7.67.

Unlike for SNPmers, there are no canonical positions that identify the
unique, unpaired \kmers\  . Several mechanisms have been proposed for sampling \kmers\  in a representative way\citep{roberts_reducing_2004,ye_exploiting_2012}. We use the more pseudo-random hash-slicing rule, already discussed above, to sample a single slice of \kmers\: those whose integer encodings are congruent to a particular slice number $s$, modulo $S$ (the hash slicing factor). We have found that on the finished human genome (results not shown), hash slicing is effectively a Poisson sampling, with sampled \kmers\  spaced according to an exponential distribution.

A caveat in applying hash slicing is that taking the remainder modulo a prime is not very pseudo-random for Mersenne primes (equal to $2^{p-1}$ for some \emph{p}), when \kmers\  are represented in base-4 encoding\citep{knuth_searching_1973}. We therefore pick a slicing factor of 11, the smallest non-Mersenne prime greater than our SNPmer sampling factor.

The resulting \kmer\  subset has 86.0 million unpaired \kmers\  and 24.0 million SNPmer pairs, each reduced as predicted, for a total factor of 9.8 reduction in \kmer\  nodes for the next step.
\subsection*{Contigging and scaffolding}
\label{sec-7-6}

Each of the sampled unique \kmers\  and SNPmer pairs is a node in the
\kmer\  graph. Nodes are connected when the corresponding \kmers\ 
appear consecutively in at least one read of the input (any
intervening \kmers\  having been skipped due to sampling or
ambiguity). The relative orientation, distance and number of
supporting reads of the \kmers\  is stored in the edge. When
conflicting distance or relative orientation is observed among
different reads for the same pair of \kmer\  nodes, all edges from both
nodes in the corresponding direction are ignored in contigging.

Robust edges for one direction from a node are defined as those
supported by a supermajority of the reads containing the \kmer: the
number of supporting reads is greater than or equal to both (1) two
plus the sum of the read counts for all other edges in that direction
and (2) twice the read count of the next-most supported edge in the
same direction. By this construction, a node has at most one robust
edge in each direction.

A mutually robust edge is defined as one that is robust going in both
directions between the two nodes it connects.

Contigs are the connected components of the subgraph consisting of
mutually robust edges. Singleton and circular contigs are reported for
diagnostic purposes, but ignored in subsequent analysis. Each 
retained ``\kmer\  contig'' of Table \ref{table:ContigStats} therefore represents
a chain of nodes for SNPmers and unique \kmers\  not shared with other
contigs.

After assembly of \kmer\  contigs, we connect them in longer structures
using the Bambus scaffolder\citep{pop_hierarchical_2004}. This
requires mapping templates (read pairs) to contigs, based on
shared \kmer\  content, and dividing the resulting graph of contigs
linked by templates into batches small enough for Bambus to
process.

Contigs are chains of \kmer\  nodes (unique or SNPmers). Since each
node can appear only once in the contigs, contigs can be represented
with a variant of the \kmer\  hash. This allows efficient tabulation of
template-to-contig overlaps in a single pass through the read
ckdata. Each overlap is detected by the number of shared \kmers\  on a
consistent diagonal (\emph{i.e.} relative sequence offset) and their span
on read and contig. Templates that overlap multiple contigs can be
used to order and orient those contigs.

We present templates linking contigs to Bambus using AMOS format\citep{treangen_next_2011} for the reads (template ends) mapped to
each contig. Reads are included only for the contig with which it
shares the most \kmers, if the span of those \{\kmers\} is $\geq k$ and
the other read-end of the template similarly qualifies in a different
contig. Bambus then infers links between contigs by matching template
identifiers shared by reads in different ``linkable contigs''.

To generate independent AMOS files for each Bambus run, the large
graph of template links is partitioned into connected components.
These components are grouped into batches for a reasonably
load-balanced parallel computation with no template links between
batches. The results are summarized as ``initial scaffolds'' in Table
\ref{table:ContigStats}.

Consensus sequence representations for \{\kmer\} contigs and scaffolds
were generated in two phases.  In the first phase, sequence spanned by
selected SNPmers and subset \{\kmers\} (see sections \ref{snpmer} and
\ref{subset} above) are joined together, separated by a number of Ns
corresponding to the number of bases not spanned by \{\kmers\} in the
subset.  In the second phase, a single pass is made through the read
data set, and stretches of Ns that are spanned by single reads are
replaced by the sequence of the read.
\subsection*{SNP phase and genotype inference \label{genotype}}
\label{sec-7-7}

Each scaffold of the \kmer\  assembly constitutes a
candidate marker for mapping.  While the depth of sequence coverage on
each member of the mapping panel is too low ( about 1X ) to directly
infer the genotype of individual members of the mapping pannel at
individual SNPs, the tight linkage between SNPs within markers means
that learning a sample's genotype at any one reveals it at the
others, effectively amplifying the sequence coverage by a factor
proportional to
the number of SNPs within the marker.  This is the same principle
exploited in genotype by sequencing (GBS) approaches to genetic
mapping in the presence of reference genomes, for example in
recombinant inbred lines of reference rice strains\citep{huang_high-throughput_2009}, and in crosses between \emph{Drosophila}
species with sequenced genomes\citep{andolfatto_multiplexed_2011}.  We use the same principle to genotype offspring in the context of a
cross between two outbred individuals, simultaneously inferring the phases of
the SNPs (\emph{i.e.} which bases appears on each of the four parental
chromosomes in the cross).    

While the data will be insufficient to infer genotypes at many
markers, all those where confident inferences can be made can be used
to build the linkage map.

For the purposes of genotype inference, a marker is treated as a
collection of $m$ SNPs (indexed in the following by $i \in \{1, 2,
... , m\}$), that have been inferred to be closely linked on the
genome via the \kmer\  assembly step.
If the four parental chromosomes are labeled $a,b$ in one parent and
$c,d$ in the other, then the genotyping problem is to infer which of
the four possible segregation states or genotypes $ac,ad,bc,bd$
describes each sample at each marker locus.  We index samples with
$j$, and denote a sample genotype by $g_j$.

We assume that markers are very small compared to a
chromosome, and ignore the possibility of a recombination event within
individual markers.
  The data used for inference of the offspring genotypes consist of the
number of reads from each barcoded sample $j$
showing each of the four
possible DNA bases $b$ at each variable SNP position $i$, which we denote ${ n_{ij}^b }$.  

If the phase $\phi_i$ of SNP $i$ were known, \emph{i.e.} which base is present in each
of the four parental chromosomes, then a choice of genotype $g_j$
implies a specific homozygous or heterozygous state $s_{ij} \in S = 
\{AA,CC, TT, GG, AC, AT, AG, CT, CG, TG \}$ for SNP $i$ in
sample $j$.
For a given phase and genotype, the likelihood function 
for a given SNP position in a given
sample is given by either a binomial (for homozygous states) or trinomial 
probability mass function  
of the read counts, base-calling error rate $\epsilon$, and the site
genotype $s_{ij}$:

\begin{equation}
%\[
 \mathbb{L}( \phi_i, g_j ) = p(n_{ij}^b | \phi_i , g_j ) = P_m(n_{ij}^b,s_{ij},\epsilon) =
\begin{cases}
    {n \choose m} \epsilon^m (1-\epsilon)^{n-m} ,& \text{if } s_{ij} \text{ homozygous} \\
    \frac{n!}{k! l! m!} \epsilon\prime^m (\frac{1-\epsilon\prime}{2})^{k+l} ,& \text{if } s_{ij} \text{ heterozygous} \\
\end{cases}
%\]
\end{equation}

where $n$ is the total number of reads at SNP $i$; $m$ is the number of observations of bases not in $\sigma_{ij}$ (\emph{i.e.} mismatches); $k$ and $l$ are the counts for each of the two bases of $\sigma_{ij}$ for heterozygous sites, and  $\epsilon\prime = 2\epsilon/3$.
\subsection*{Likelihood maximization}
\label{sec-7-8}

Searching for an optimal choice of SNP phases $\phi_i$ and sample genotypes $g_j$ is made difficult by the exponential size of the search space: for segregating bi-allelic SNP sites there are 14 possible phases to consider at each SNP site, so for a mapping panel of only 20 siblings and a marker containing only 10 SNPs, there are $4^{20} \times 14^{10} > 3 \times 10^{23}$ combinations to consider.  In simulation tests, we found that a variant of expectation maximization (EM), an iterative likelihood maximization method can accurately infer a large proportion of marker genotypes.  

To initialize the iteration, the parental samples and a randomly selected offspring are, without loss of generality, assigned genotypes $(a,b)$, $(c,d)$ and $(a,c)$.  At each step, we calculate the conditional probability distributions over the possible SNP phases $p(\phi_i)$ given the genotype assignments according to:

\begin{equation}
p^{(t)}(\phi_i) = p(\phi_i|\mathbf{g}^{(t)}) = \prod_{\sigma \in S}
p(n_{i\sigma}^b|\phi_i,\mathbf{g}^{(t)}) / ( 
\sum_k\prod_{\tau \in S} p(n_{k\tau}^b|\phi_k,\mathbf{g}^{(t)}) )
\end{equation}

where we have labeled the chosen values for the genotypes $g_j$ at iteration $t$ collectively by $\mathbf{g}^{(t)}$. $n_{i\sigma}^b$ is the combined total number of observations of base $b$ at polymorphic SNP $i$ for all samples included at iteration step $t$ which have genotype $s(\phi_i,g_j) = \sigma$.

On each iteration until all samples have been included, a randomly selected sample is added to the set after calculating
$p^{(t)}(\phi_i)$. Then the next set of genotype assignments $\mathbf{g}^{(t+1)}$ are determined by
choosing those that maximize the expected value of the log likelihood:

\begin{equation}
\mathbb{E}_{\phi|n,g_j} [ \log L(g_j;n,\phi) ]= \sum_i p(\phi_i) \log p(n_{ij}^b|g_j,\phi_i)  \label{eqExp}
\end{equation}

These steps are repeated until genotypes are being selected for all
samples, and the expected log likelihood stops increasing.  At the end
of the iteration, the likelihood-maximizing genotypes are reported,
along with the log likelihood difference between the best and second
best choice of genotype for each sample, which provides an indicator
of the confidence in genotype call.
To gauge  convergence, this procedure is repeated 5 times for each marker, with different random choices of initial conditions.  Markers which do not identify the same ML genotype multiple times in independent runs are not included among the high confidence genotype calls.
\subsection*{Error model calibration \label{errormodel}}
\label{sec-7-9}

The sequence of 14 \emph{Ciona intestinalis} autosomes were downloaded from Ensembl (www.ensembl.org).
These 14 chromosomes were used as the template in our genome
simulation. Based on their sequence length, We used a markovian
coalescent simulator \texttt{macs} \citep{chen_fast_2009} to generate four
haploid samples drawn from a population under neutral Wright-Fisher
model with population mutation rate of 0.012 and population
recombination rate of 0.0085. Using the \emph{C. intestinalis} genome as
the reference sequence, two diploid parental genomes were constructed
based on the \texttt{macs} output with realistic SNP and Indel models
inferred by several previous studies on the Ciona
genome\citep{dehal_draft_2002,haubold_mlrho_2010,small_extreme_2007}. We
wrote a \texttt{perl} script to simulate the genomes of offspring generated
by the cross of the two simulated parents. The software
package \begin{tt}dwgsim\end{tt}\citep{homer_dwgsim_2012} was used to
generate Illumina paired-end reads based on our simulated genomes of
both parents and offsprings, with the coverage of 20X and 5X
respectively.

To estimate the frequency of incorrect genotype calls as a function of
the log likelihood difference between the called and alternative
genotype, including contributions from uncertainty in SNP-mer
identification, assembly, and sampling noise, we carried out a
simulation of the \kmer\  assembly and genotype inference protocols
Among high-confidence genotype calls, the observed error frequency was
a function of call confidence score was well-fit by a sum of two
stretched exponential functions, allowing assignment of
error probabilities to individual genotype calls.
\subsection*{Linkage group construction}
\label{sec-7-10}

We use the linkage p-value $p_{ab}$ between pairs of map bins $a$ and $b$ defined as the minimum over the four 
possible relabelings $r$ of the maternal and paternal chromosomes of the Binomial p-value for the number of matching genotypes:

\begin{equation}
p_{ab} = \min_r \left[1-\sum_i^{ m_r-1 } {n\choose i} \frac{1}{2^n} \right]
\end{equation}

where $n$ is the total number of sample genotype calls (68 in the present case, or 34 in each parent) and $m_r$ is the number of matching genotypes under relabeling $r$.  

We identified map bins with segregation patterns indicating either inconsistent placement in the maternal and paternal maps or genotyping error with a double threshold procedure as follows:

\begin{enumerate}
\item Map bins were partitioned into linkage groups by single linkage clustering at a threshold of $p_{ab} < p_1$.
\item Within each partition, map bins which formed articulation points\citep[\begin{em}i.e.\end{em} nodes which, if removed, would cause the linkage group to fall apart into two disconnected subgraphs;][]{hopcroft_algorithm_1973} in the graph of $p_{ab} < p_2$, where $p_2 > p_1$.
\end{enumerate}

This procedure identifies map bins which alone account for the merging of what would otherwise be two distinct partitions.  We used the following pairs of thresholds $p_1,p_2$ to identify a total of 20 map bins for exclusion from the map:  
10$^{\mathrm{-7}}$,10$^{\mathrm{-6}}$; 
10$^{\mathrm{-8}}$,10$^{\mathrm{-7}}$; 
10$^{\mathrm{-9}}$,10$^{\mathrm{-8}}$.
The remaining markers form locally consistent linkage groups in which all linkages defined at threshold $p_1$ are corroborated by multiple linkages at $p_2$, for the above values of $p_1$ and $p_2$.
\subsection*{Marker ordering}
\label{sec-7-11}

Markers were ordered within each linkage group using the following
protocol.  Within each linkage group a consistent labeling of the four parental chromosomes was achieved by constructing a graph $G$ in which nodes correspond to map bins and edges are weighted by linkage p-value $p_{ab}$ (as defined above).  The local chromosome labels are updated at each map bin as it is reached in a traversal of the minimum spanning tree of $G$ to the labeling $r$ that maximizes $p_{ab}$ along the incident of $G$ used in the traversal.
Markers within each linkage group were clustered by hierarchical clustering (marker-marker distance metric: cosine of the angle between the vectors of recombination distances to the other map bins; distance updating method: average linkage) into a binary tree data structure with leaves representing map bins.  A node in the right subtree of the root node was rotated, interchanging its left and right subtrees if its left subtree was not already closer (in average recombination distance) to the markers of the left subtree of the global root; and similarly for nodes in the left subtree of the root.  An in-order traversal of the tree generates an ordering of the markers.  Finally three reversals of the order of markers in segments of the map were added based on visual inspection of the recombination distance matrix.
In the final marker ordering, 51\% of adjacent map bin pairs are separated by a single recombination event in the cross, and 94\% are separated by three or fewer recombinants in each parent.
\subsection*{Placement of markers on the map \label{placement}}
\label{sec-7-12}

To anchor additional markers to the map, we computed the $p_{ab}$ (see
above) between marker $a$ to be placed on the map and each map bin
$b$.  Marker $a$ is anchored to the map at the position of the bin $b$
which minimizes $p_{ab}$ if $p_{ab} < 10^{-6}$.
\subsection*{SNP density estimation \label{samtools}}
\label{sec-7-13}

Illumina reads were mapped to the assembled scaffold sequences with \texttt{stampy}\citep{lunter_stampy:_2011} using default settings.  For a
sample of 9,228 scaffolds with lengths ranging from 5.0-5.5 kb,
sequence variants were called with \texttt{SAMtools}\citep{li_statistical_2011}
using a variant quality score threshold of 50, and ignoring indel
positions.

A SNP density of 0.76 \% in four haplotypes which
corresponds to a predicted rate of pairwise sequence differences per
site of $\theta$ = 0.0042 under the finite sites model of mutation and
the neutral coalescent model of the relationships among sampled
alleles\citep{yang_statistical_1996}. 
\subsection*{Estimation of local recombination rate}
\label{sec-7-14}

To estimate the local recombination rate for each map bin, we computed
the linear regression of map distance in number of markers on physical
distance using up to 10 neighboring map bins in each direction along
the map (or fewer for bins within 10 map bins of the end of the
linkage group).  Map distance was calculated from recombination
fraction using Haldane's map distance $-\frac{1}{2}log(1-2r)$ \citep{haldane_combination_1919}. 
\subsection*{Ancestral linkage group conservation}
\label{sec-7-15}

To compare the genome organization in \emph{L. polyphemus} to the ancestral
metazoan ALGs, we used the reciprocal best blast hit (RBH) orthology
criterion in an alignment of the \emph{Ixodes scapularis} predicted
proteins\citep{vectorbase_ixodes_2008} to the consensus sequences for
the marker scaffolds.  \emph{L. polyphemus}
scaffolds with RBH of e-value $<= 10^{ \mathrm{-6}} $ were assigned to the same
ancestral bilaterian gene orthology group as their \emph{I. scapularis} ortholog,
and thereby with human genes.
Regions of the map were tested for enrichment in genes from particular
ancestral linkage groups with Fisher's Exact Test, and breakpoints in
ancestral linkage group composition were identified using a hidden
Markov model, as previously
described\cite{putnam_sea_2007,putnam_amphioxus_2008}.
\subsection*{Homeobox gene modeling \label{modeling}}
\label{sec-7-16}

We identified 155 marker scaffolds with a tblastx alignment of 
e-value $ < 10^{\mathrm{-6} } $ to a set of chelicerate homeobox 
gene sequences downloaded from Genbank using the 
NCBI online query interface (genbank accessions 
AF071402.1, AF071403.1, AF071405.1, AF071406.1, AF071407.1, AF085352.1, AF151986.1, AF151987.1, AF151988.1, AF151989.1, AF151990.1, AF151991.1, AF151992.1, AF151993.1, AF151994.1, AF151995.1, AF151996.1, AF151997.1, AF151998.1, AF151999.1, AF152000.1, AF237818.1, AJ005643.1, AJ007431.1, AJ007432.1, AJ007433.1, AJ007434.1, AJ007435.1, AJ007436.1, AJ007437.1, AM419029.1, AM419030.1, AM419031.1, AM419032.1, DQ315728.1, DQ315729.1, DQ315730.1, DQ315731.1, DQ315732.1, DQ315733.1, DQ315734.1, DQ315735.1, DQ315736.1, DQ315737.1, DQ315738.1, DQ315739.1, DQ315740.1, DQ315741.1, DQ315742.1, DQ315743.1, DQ315744.1, EU870887.1, EU870888.1, EU870889.1, HE608680.1, HE608681.1, HE608682.1, HE805493.1, HE805494.1, HE805495.1, HE805496.1, HE805497.1, HE805498.1, HE805499.1, HE805500.1, HE805501.1, HE805502.1, S70005.1, S70006.1, S70008.1, and S70010.1).  The reads of each marker (those with best \texttt{stampy}\citep{lunter_stampy:_2011} alignment to the scaffold) were reassembled with \texttt{PHRAP}\citep{gordon_consed:_1998}, with default parameters.  The resulting contigs were aligned to a collection of homeobox-containing protein sequences (genbank accessions NP 001034497.1, NP 001034510.1, AAL71874.1, NP 001034505.1, NP 001036813.1, CAA66399.1, NP 001107762.1, NP 001107807.1, EEZ99256.1, NP 001034519.1, NP 476954.1, NP 032840.1, NP 031699.2, AAI37770.1, EEN68949.1, NP 523700.2, NP 001034511.2, AAK16421.1, and AAK16422.1) with exonerate\citep{slater_automated_2005} in protein-to-genome mode.  For each contig, the amino acid sequence predicted by the highest-scoring exonerate alignment was used in subsequent phylogenetic analysis, resulting in 104 putative homeobox-containing markers ranging in length from 18 to 147 amino acids.  
\subsection*{Phylogenetic analysis of homeobox genes}
\label{sec-7-17}

\begin{figure}[htb]
\centering
\includegraphics[height=8in, angle=0]{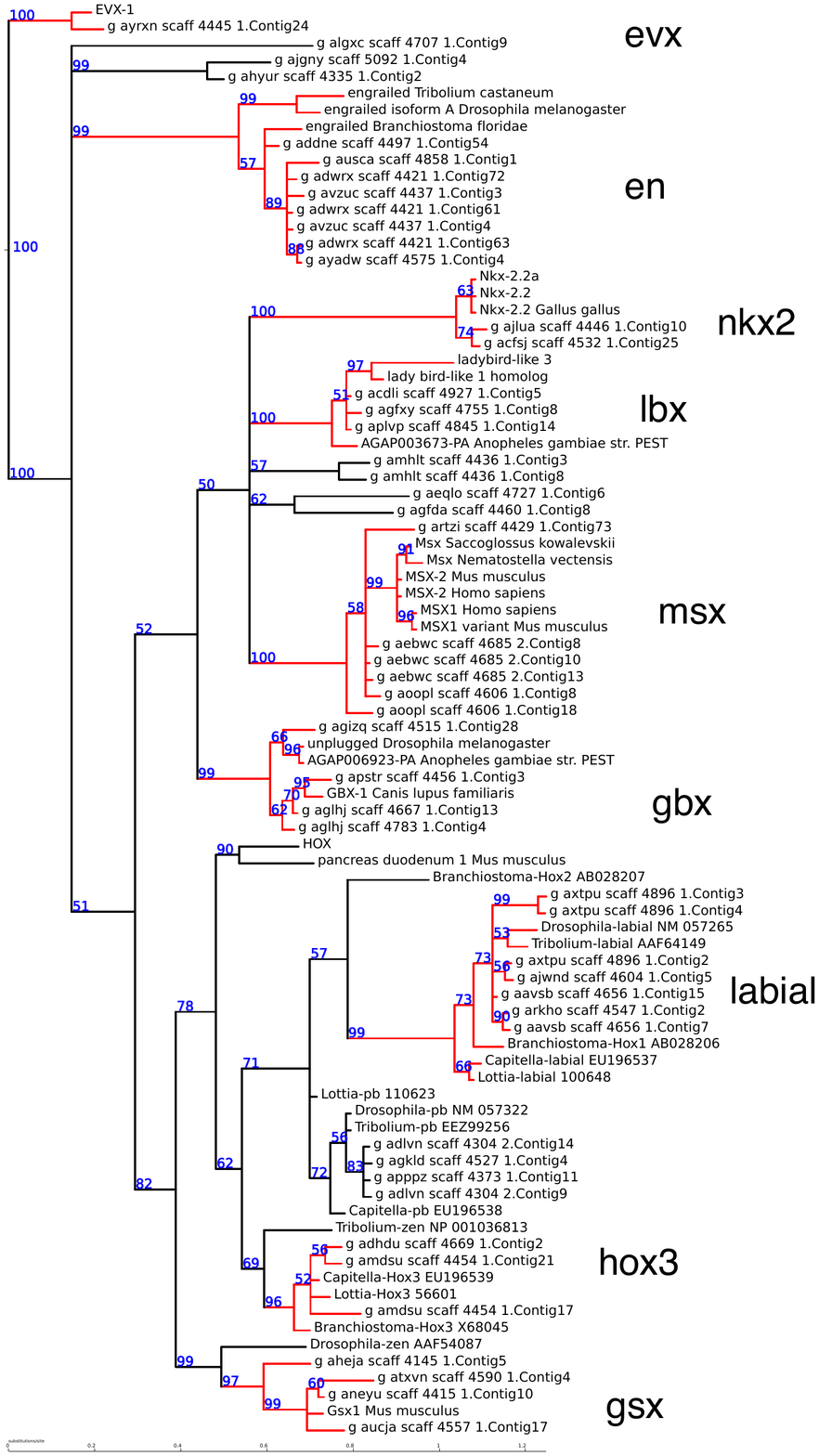}
\caption{\label{fig:hoxtree1}Unrooted phylogenetic tree of homeobox sequences (part 1).  Nodes are labeled with Bayesian posterior probabilities.  Highly supported partitions used to classify \emph{L. polyphemus} sequences are drawn in read, with the abbreviation for the class shown in large letters.  \emph{L. polyphemus} homeobox sequences not grouped into one of these highly supported partitions are assigned to class ``?''.  For ease of display, a large subtree consisting of HOX and parahox genes has been pruned at the position labeled ``HOX'', and is shown in Figure \ref{fig:hoxtree2}.}
\end{figure}

\begin{figure}[htb]
\centering
\includegraphics[height=8in, angle=0]{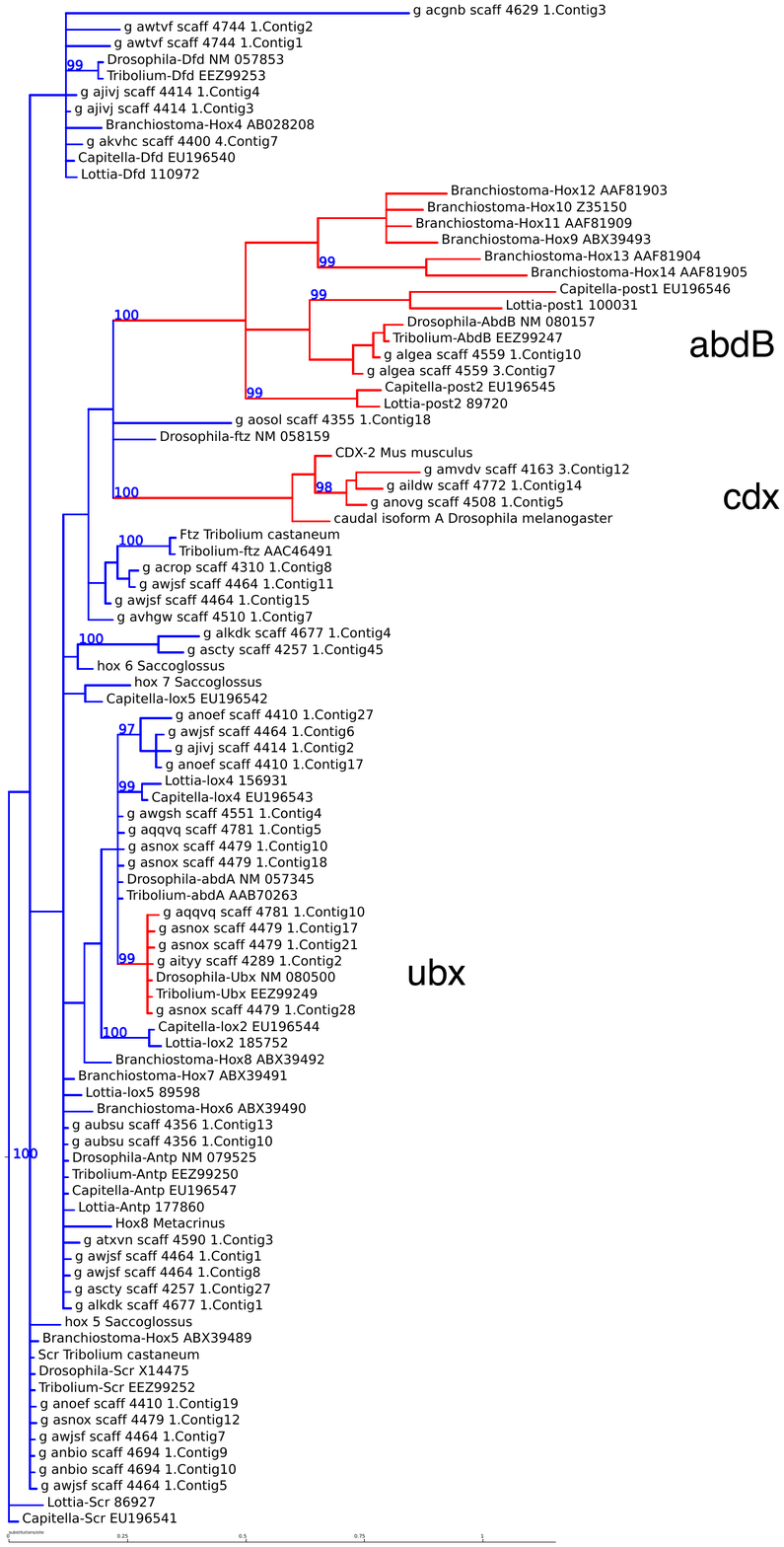}
\caption{\label{fig:hoxtree2}Phylogenetic tree of homeobox sequences, part 2.  The rooted subtree pruned from the tree in Figure \ref{fig:hoxtree1}.  Nodes are labeled with Bayesian posterior probabilities.  Highly supported partitions used to classify \emph{L. polyphemus} sequences are drawn in read, with the abbreviation for the class shown in large letters.  \emph{L. polyphemus} homeobox sequences not grouped into one of these highly supported partitions are assigned to class ``hox?''.}
\end{figure}

A multiple sequence alignment of the predicted homeobox sequences
combined with a collection of representative sequences from various
classes of homeobox genes was constructed with muscle3.8.31\citep{edgar_muscle:_2004} using default settings.  The resulting
alignment was trimmed to a 63 amino acid segment spanning the
conserved homeodomain, and sequences with more than 50\% gaps were
removed, leaving 93 predicted \emph{L. polyphemus} homeobox genes in the
analysis.  Bayesian phylogenetic analysis was carried out on the
resulting 178 taxon, 63 amino acid character matrix (Supplementary
File HombeoboxTable.html) using MrBayes
v3.2.1\citep{ronquist_mrbayes_2003} using a mixed model of amino acid
substituions, gamma-distributed rate variation among sites with fixed
shape parameter $\alpha=1.0$, alignment gaps treated as missing data,
2,000,000 Monte Carlo steps, two independent runs with four Monte
Carlo chains, and the initial 25\% of sampled trees were discarded as
``burn-in''.  Monte Carlo appeared to reach convergence, with an average
standard deviation of the split frequencies of 0.022.
The majority-rule consensus of the sampled trees is shown in Figures
\ref{fig:hoxtree1} and \ref{fig:hoxtree2}, and well-supported gene
clades (posterior probability greater than 0.95) were used to group
the predicted \emph{L. polyphemus} genes into classes.  The table in
Supplementary File HombeoboxTable.html lists the reassembled marker
contigs, their inferred hox gene class, and maximum likelihood map
positions.  Predicted genes were anchored to the map as 
described above.

\subsection*{Genomic distribution of paralogs}
\label{sec-7-18}

We identified 2716 pairs of \emph{Limulus} markers that can both be placed
on the map and have their best translated alignment to the same
\emph{I. scapularis} gene. (\emph{I. scapularis} genes with more than five
best-hit markers were excluded from seeding such pairs.)  To estimate
the synonymous sequence divergence between pairs of candidate
\emph{L. polyphemus} paralogous gene pairs and \emph{L. polyphemus} genes and
their \emph{T. tridentatus} orthologs, we constructed codon alignments of
predicted coding sequence for estimation of synonymous sequence
divergence.  Conserved clusters of paralogs were identified using a
variant of the ``max-gap'' criterion\citep{mclysaght_extensive_2002} in
which two genes are placed in the same cluster if they and their
paralogs lie within threshold distance.  In the analysis presented
here, the distance threshold used was 500 markers.

\newpage
\subsection*{K$_a$ and K$_s$ estimation for paralogs and \emph{T. tridentatus} orthologs \label{paralogs}}
\label{sec-7-19}

\begin{figure}[htb]
\centering
\includegraphics[height=5in]{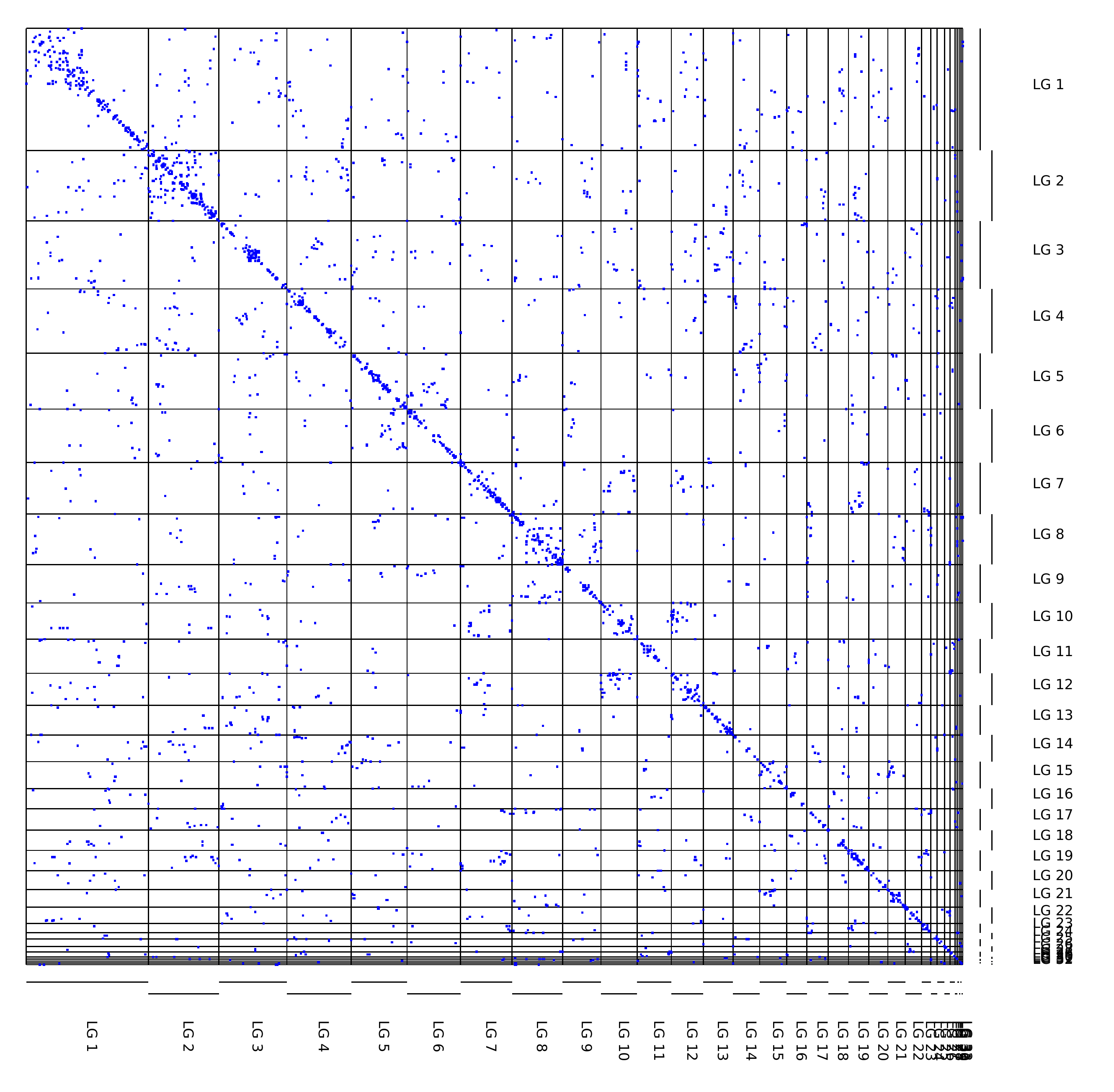}
\caption{\label{fig:paralogdots}Genomic distribution of candidate paralogs}
\end{figure}

Figure \ref{fig:paralogdots} shows the distribution across the map of pairs of candidate paralogs.

To estimate the synonymous sequence divergence between pairs of candidate \emph{L. polyphemus} paralogous gene pairs, and between \emph{L. polyphemus} genes and their \emph{T. tridentatus} orthologs, we followed the following protocol.

\begin{enumerate}
\item Reassemble reads from each marker with \texttt{PHRAP}\citep{gordon_consed:_1998}, and create a
    predicted coding sequence using exonerate, as described for the
    annotation of homeobox gene models (see \ref{modeling} above).
\item Combine the exonerate alignments of codons to amino acids to
    create an alignment of codons for either a pair of
    \emph{L. polyphemus} sequences, or for a \emph{L. polyphemus} -    \emph{T. tridentatus} sequence pair.
\item Use the method of Yang and Nielsen\citep{yang_estimating_2000} to estimate the synonymous and
    non-synonymous substitution rates K$_a$ and K$_s$, as implemented in
    the KaKsCalculator package\citep{zhang_kaks_2006}.
\item Discard estimates based on fewer than 30 sites (30 synonymous sites for estimates of $K_s$, non-synonymous sites for $K_a$).
\end{enumerate}

GenBank accessions for \emph{Tachypleus tridentatus} mRNA clones: JQ966943, AB353281, AB353280, HM156111, HQ221882, HQ221883, HQ221881, HQ386702, HM852953, TATTPP, TATPROCLOT, FN582225, FN582226, AF467804, AF227150, GQ260127, AF264067, AF264068, AB353279, AB005542, TATLICI, TATTGL, TATCFGB, TATLFC1, TATLFC2, AB201713, TATCFGA, TATLICI2, CS423581, CS423579, AB028144, AB201778, AB201776, AB201774, AB201772, AB201770, AB201768, AB201766, AB201779, AB201777, AB201775, AB201773, AB201771, AB201769, AB201767, AB201765, AB105059, AB002814, AX763473, TATCFBP, AB076186, AB076185, X04192, TATHCLL, AB037394, AB019116, AB019114, AB019112, AB019110, AB019108, AB019106, AB019104, AB019102, AB019100, AB019098, AB019096, AB019117, AB019115, AB019113, AB019111, AB019109, AB019107, AB019105, AB019103, AB019101, AB019099, AB019097, AB023783, AB024738, AB024739, AB024737, AB017484, D87214, D85756, D85341.

\begin{figure}[htb]
\centering
\includegraphics[height=5in]{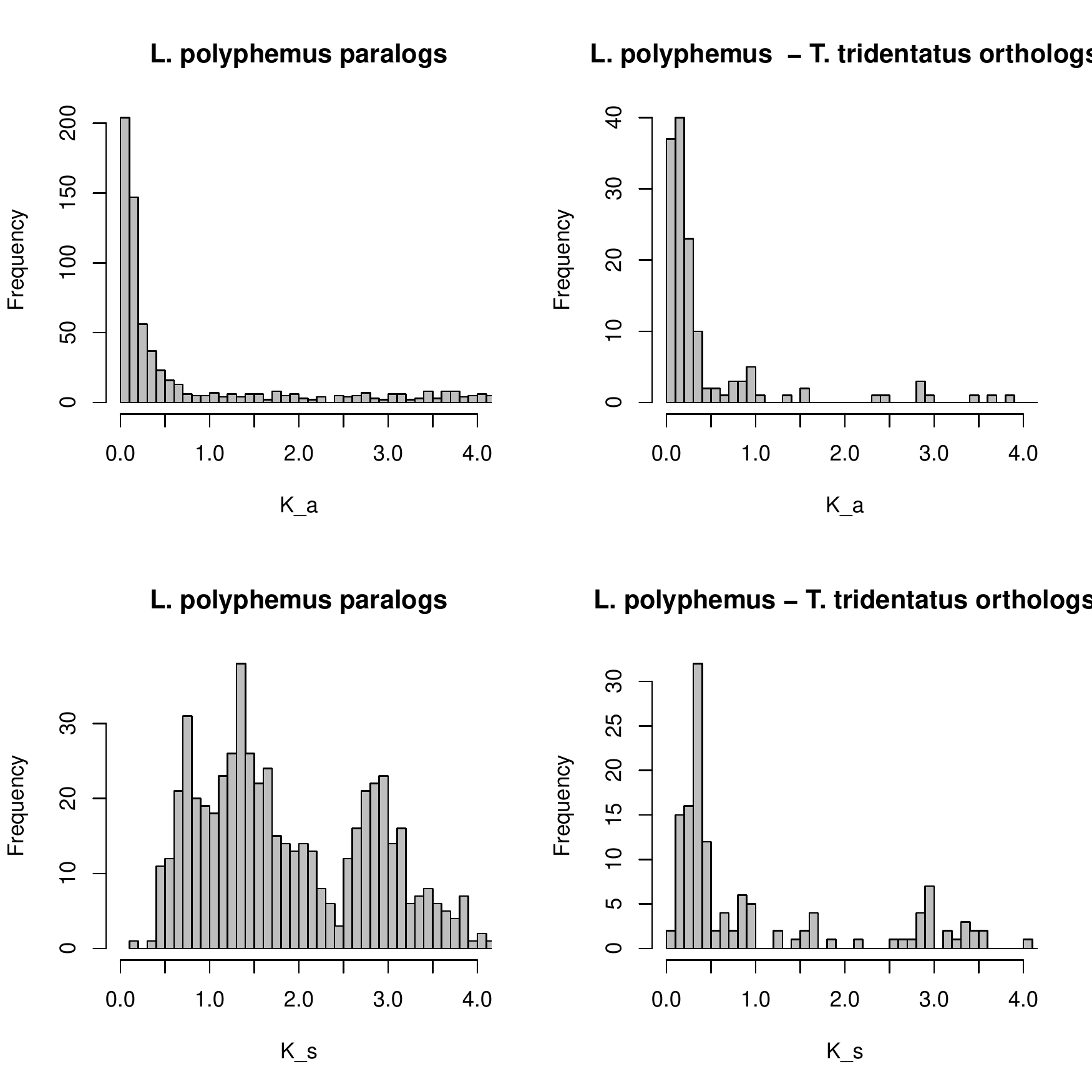}
\caption{\label{fig:ks}Distribution of synonymous and non-synonymous sequence divergence rates for pairs \emph{L. polyphemus} paralogs and \emph{L. polyphemus} - \emph{T. tridentatus} orthologs.}
\end{figure}

\begin{figure}[htb]
\centering
\includegraphics[height=3in]{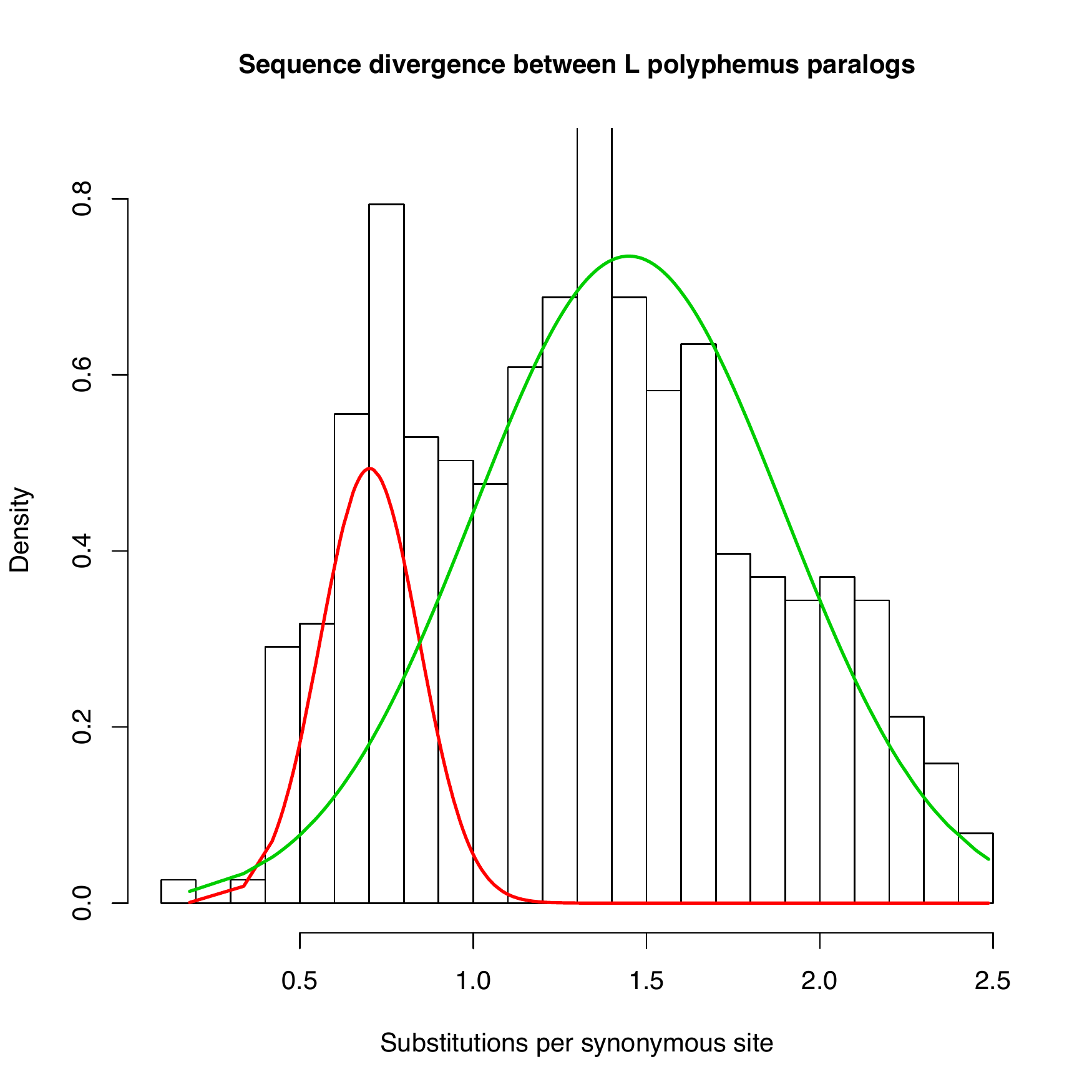}
\caption{\label{fig:mixture}Two component mixture model fit to the $K_s$ peak on the range $0<=K_s<=2.5$.}
\end{figure}

Figure \ref{fig:ks} shows the distributions of K$_a$ and K$_s$ for paralogs and  \emph{T. tridentatus} orthologs.  To estimate the number and age of peaks in the un-saturated range\citep{vanneste_inference_2013} of the K$_s$ distribution (and of putative WGD events), 
we fit a series of univariate normal mixture models, with 1, 2, 3, and
4 components to the paralog $K_s$ distribution in the 
range $0 < K_s < 2.5$ and selected 
the best model on the basis of Bayesian Information Criterion (Table \ref{table:ks}).
The best model had two components, with means at 0.7 and 1.45 substitutions per site.  The position of the peak at lowest K$_s$ was not sensitive to the addition of more mixture components.
Figure \ref{fig:mixture} shows a comparison of the distribution and the components of the best-fitting model.
Gaussian mixture models were estimated in \texttt{R} with \texttt{mixtools}\citep{benaglia_mixtools:_2009}.

\begin{table}[htb]
\caption{Mixture model fits to K$_s$ distribution.  N is the number of mixture components, k the number model parameters, ln(L) the log likelihood of the data under the best fit model, BIC the Bayesian information criterion, AIC the Akaike information criterion.} \label{table:ks}
\begin{center}
\begin{tabular}{rrrrrl}
 N  &   k  &    ln(L)  &              BIC  &     AIC  &  mixture components                                                                     \\
\hline
 1  &   2  &  -273.93  &           559.74  &  551.87  &  1.32 \textpm{} 0.50                                                                    \\
 2  &   5  &  -259.32  &  \textbf{548.31}  &  528.64  &  0.70 \textpm{} 0.14 ; 1.45 \textpm{} 0.45                                              \\
 3  &   8  &  -253.36  &           554.19  &  522.71  &  0.71 \textpm{} 0.17 ; 1.34 \textpm{} 0.29 ; 2.09 \textpm{} 0.19                        \\
 4  &  11  &  -251.41  &           568.11  &  524.82  &  0.74 \textpm{} 0.18 ; 1.34 \textpm{} 0.20 ; 1.70 \textpm{} 0.04 ; 2.02 \textpm{} 0.22  \\
\end{tabular}
\end{center}
\end{table}

\section*{Data Access}
\label{sec-8}

The raw sequencing reads are currently being submitted through the NCBI SRA and are
accessible via NCBI BioProject accession PRJNA187356.  
\section*{Acknowledgements}
\label{sec-9}

This research was supported by the National Science Foundation
(EF-0850294 and IOB 06-41750), the Beckman Young Investigator Program,
the University of Florida Division of Sponsored Research, the
Department of Biology, and the UF Marine Laboratory at Seahorse Key.
\section*{Author contributions}
\label{sec-10}

NHP conceived and led the project.  All authors wrote  the paper.  
PH, NHP and J-XY wrote software.  NHP, PH, J-XY and CWN
 carried out sequence analysis.  JB collected and raised samples.  CWN
 extracted genomic DNA and created the libraries for sequencing. 
\section*{Discosure Declaration}
\label{sec-11}

The authors declare that they have no conflicts of interest.
\section*{Figures}
\label{sec-12}
\section*{}

\bibliographystyle{plos2009}
\bibliography{putnamlab5}
\subsection*{}

\end{document}